\documentclass[twocolumn,secnumarabic,amssymb, nobibnotes, aps, prd]{revtex4-2}

\usepackage{hyperref}       
\usepackage{amsmath}
\usepackage{adjustbox}
\usepackage{graphicx}
\usepackage{booktabs}       
\usepackage{amsfonts}       
\usepackage{nicefrac}       
\usepackage{microtype}
\usepackage{cleveref}
\usepackage{xcolor}
\usepackage{multirow}
\usepackage{makecell}
\usepackage{array}

\begin{document}

\title{Radial Oscillations in Neutron Stars with Delta Baryons}%
\author{Ishfaq A. Rather$^1$}
\email{ishfaq.rather@tecnico.ulisboa.pt}
\author{Kauan D. Marquez$^2$}
\email{marquezkauan@gmail.com}
\author{Grigoris Panotopoulos$^3$}
\email{grigorios.panotopoulos@ufrontera.cl}
\author{Il{\'i}dio Lopes$^1$}
\email{ilidio.lopes@tecnico.ulisboa.pt}
\affiliation{$^1$Centro de Astrof{\'i}sica e Gravita{\c c}{\~a}o-CENTRA, Instituto Superior T{\'e}cnico,
Universidade de Lisboa, 1049-001 Lisboa, Portugal;}
\affiliation{$^2$Departamento de F{\'i}sica - CFM, Universidade Federal de Santa Catarina, Florian{\'o}polis/SC, CEP 88.040-900, Brazil;}
\affiliation{$^3$Departamento de Ciencias F{\'i}sicas, Universidad de la Frontera, Casilla 54-D, 4811186 Temuco, Chile.}
\begin{abstract}
 We investigate the effect of $\Delta$ baryons on the radial oscillations of neutron and hyperon stars, employing a density-dependent relativistic mean-field model. The spin-$3/2$ baryons are described by the Rarita-Schwinger Lagrangian density. The baryon-meson coupling constants for the spin-3/2 decuplet and the spin-1/2 baryonic octet are calculated using a unified approach relying on the fact that the Yukawa couplings present in the Lagrangian density of the mean-field models must be invariant under the SU(3) and SU(6) group transformations. We calculate the 20 lowest
eigenfrequencies and corresponding oscillation functions of $\Delta$-inclusive nuclear (N+$\Delta$) and hyperonic matter (N+H+$\Delta$) by solving the Sturm-Liouville boundary value problem and also verifying its validity. We see that the lowest mode frequencies for N+$\Delta$ and N+H EoSs are higher as compared to the pure nucleonic matter because of the deltas and hyperons present. Furthermore, the separation between consecutive modes increases with the addition of hyperons and $\Delta$s.
\end{abstract}
\maketitle

\section{Introduction}

The densest observed stars in the universe, neutron stars (NSs), serve as natural laboratories for the investigation of cold dense nuclear matter. The equation of state (EoS) of nuclear matter is the decisive factor that, theoretically, governs the structure and properties of NSs. To ensure the stability of nuclear matter inside NSs, they contain some amount of protons as well, apart from the neutrons. Because of the strong interaction's nonperturbative nature, we still know relatively little about the EoS of dense nuclear matter, especially at densities considerably higher than the nuclear saturation density ($\rho_0$), where exotic degrees of freedom are likely to exist. Nearly all theoretical descriptions of these objects encompass the entire spin-1/2 baryon octet i.e, nucleons and hyperons \cite{Glendenning:1997wn}. These investigations gave rise to the widely discussed ``hyperon puzzle". Hyperons soften the EoS, leading to a lower
maximum mass on the mass-radius curve of stars. \cite{2017hspp.confj1002B}. 

Within the relativistic mean-field approach, Glendenning \cite{1985ApJ...293..470G} considered several exotic degrees of freedom such as hyperons, kaons, and delta baryons ($\Delta$) in the NS matter. With the coupling parameters chosen, he found that the $\Delta$ baryons could be present only at densities $\approx$ 10 $\rho_0$ inside the NSs. However, recent studies have shown that with the proper couplings between $\Delta$ baryons and nucleons constrained by several experiment measurements, they might be present inside the NSs \cite{PhysRevC.67.038801, LI2018234, RADUTA2021136070, PhysRevC.106.055801, PhysRevD.107.036011, PhysRevD.102.063008} and that they could in fact make up a large fraction of the baryons in NS matter and have a significant effect on the properties of NSs. Also, since $\Delta$ baryons are approximately 30\% heavier than the nucleons ($m_{\Delta}$ = 1232 MeV) and even lighter than the heaviest spin-1/2 baryons of the octet ($\Xi$ hyperons), it is reasonable to expect the $\Delta$ baryons to exist inside NSs at almost the same density range as the hyperons ($\approx$ 2-3$\rho_0$). 

We have so far been capable of investigating the properties of dense matter under extremely difficult circumstances thanks to the recent accomplishment of gravitational wave (GW) detection by LIGO and Virgo Collaborations (LVC) of a binary neutron star (BNS) merger GW170817 event \cite{PhysRevLett.119.161101, PhysRevLett.121.161101}. GWs produced by the coexistence of BNS merger events provide enough information to significantly constrain the EoS and the internal composition of NSs \cite{PhysRevLett.119.161101, PhysRevLett.121.161101, PhysRevLett.120.261103, PhysRevLett.120.172702, Capano2020}. The oscillating NSs also emit GWs with several frequency modes that can be used to investigate the internal constituents and hence various properties of the star \cite{PhysRevLett.77.4134, 10.1046/j.1365-8711.1998.01840.x}. 

 Following their formation in the supernovae, oscillating NSs emit a range of frequencies depending on the restoring force and there are numerous mechanisms that could be the diverse causes of these oscillations \cite{PhysRevLett.116.181101, PhysRevLett.108.011102, Chirenti_2017}. Overall, oscillations can be divided into two categories: radial and non-radial.  In a pioneering work, Chandrasekhar  \cite{PhysRevLett.12.114, 1964ApJ...140..417C} investigated the radial oscillations of stellar models. Importantly, radial oscillation characteristics can reveal details regarding the stability and EoS of compact stars. As radial oscillations cannot produce GWs on their own, their detection is rather difficult. They are linked to non-radial oscillations, which amplify GWs and improve the likelihood of detecting them \cite{PhysRevD.73.084010,PhysRevD.75.084038}. However, \citet{Chirenti_2019} observed that in the post-merger event of BNS, a hyper-massive NS is created along with the emission of a short gamma-ray burst (SGRB), which could be influenced by radial oscillations. The high-frequency oscillations of the hyper-massive NS in the range of 1-4 kHz could be observed.

 Similar to different families of modes arising from different physical origins as described in Ref. \cite{kokkostas}, the radial oscillation modes can also be categorized into two families that are largely independent of one another. One family resides mostly in the neutron star's high-density core, whereas the other resides primarily in its low-density envelope \cite{1997A&A...325..217G}. The significant shift in the matter's stiffness at the neutron drip point causes a ``wall'' in the adiabatic index that separates the two regions. Given that it is related to the neutron drip point, which is a part of the low-pressure regime and is the same for all EoSs, this wall effect is present for any realistic EoS. 
 
 We investigate various radial oscillations of NSs with different matter compositions in this work. Several studies on the investigation of various radial oscillations of NSs with different exotic phases such as dark matter and deconfined quark matter have already been carried out \cite{10.1093/mnras/stac2622, kokkostas, https://doi.org/10.48550/arxiv.2205.02076, PhysRevD.101.063025, PhysRevD.98.083001, https://doi.org/10.48550/arxiv.2211.12808}. But the radial oscillation of NSs with $\Delta$ baryons ($\Delta$-inclusive nucleonic stars) and hyperon stars with $\Delta$ baryons ($\Delta$-inclusive hyperonic stars) is being studied for the first time.

  The neutron star EoS at supra-nuclear densities has been constructed using a variety of models with a range of parameterizations. The saturation properties of highly dense nuclear matter have been extensively studied using density functional theories (DFT), in which the nucleon-nucleon interaction is effectively defined by fitting ground state properties of finite nuclei \cite{PhysRevC.5.626, SHEN1998435, PhysRevC.65.035802,refId0, PhysRevC.89.045807, PhysRevC.90.045802}.    
  From many-body theories, the nuclear matter EoS at saturation density is well constrained. The properties of neutron stars are interpreted by extrapolating these EoSs to densities several times the nuclear saturation density. 
   
    The density-dependent relativistic mean-field (DD-RMF) model \cite{PhysRevLett.68.3408} is a widely used and successful model with the advantage that the self- and cross-coupling of various mesons in the RMF model are replaced by the density-dependent nucleon-meson coupling constants.
    The results produced by the density-dependent coupling constants are comparable to those of other models and allow for consistent measurement of NS properties. It takes into account the Dirac-Brueckner model's characteristics and uses microscopic interactions at varying densities as an input.
  DD-RMF parameter sets such as DD-ME1 \cite{PhysRevC.66.024306}, DD-ME2 \cite{PhysRevC.71.024312}, and DD-MEX  \cite{TANINAH2020135065} generate a very stiff EoS and hence predict a massive NS with a maximum mass in the range of 2.3-2.5$M_{\odot}$ \cite{Rather_2020, PhysRevC.103.055814}.  
     
Our work is organized as follows: in Section (\ref{sec:eos}), the EoS for the DD-RMF model along with the addition of $\Delta$ baryons and the couplings used is discussed.  The Sturm-Liouville
eigenvalue equations for the internal structure and radial
oscillations of NSs are introduced in Section (\ref{radial}). In Section (\ref{results}), the EoS and the Mass-Radius profile for different compositions of the matter are discussed in Section (\ref{mr}). Section (\ref{profile}) describes the numerical results obtained for NSs and hyperon stars with $\Delta$ baryons. The summary and concluding remarks are finally given in Section (\ref{summary}). \par 


\section{Theoretical framework and formalism}

\subsection{Equation-of-state}\label{sec:eos}

In this study, the hadronic matter composing the neutron stars is described  within a relativistic mean-field approach with density-dependent couplings (DD-RMF). 
This type of model is shown to be very consistent in the description of nuclear matter experimental properties and also holds  when astrophysical constraints are imposed  \cite{dutra2014,lourencco2019,malik2022}.
The interaction is described considering nucleons (and other hadrons) interacting through the exchange of virtual mesons, and the DD-RMF model adopted here considers the scalar meson $\sigma$, the vector mesons $\omega$ and $\phi$ (that carries hidden strangeness), isoscalars,  and the isovector-vector meson $\vec\rho$.
The Lagrangian density is the basic ansatz of any RMF theory and contains the contributions from free baryons and mesons together with  the terms describing the interaction between them.

The lagrangian of the relativistic model in the mean field approximation used to describe the hadronic interactions here is given by
\begin{widetext}
\begin{align}
     \mathcal{L}_{\rm RMF}=  {}& 
     \sum_{b\in H}  \bar \psi_b \Big[  i \gamma^\mu\partial_\mu - \gamma^0  \big(g_{\omega b} \omega_0  +  g_{\phi b} \phi_0+ g_{\rho b} I_{3b} \rho_{03}  \big)- \left( m_b- g_{\sigma b} \sigma_0 \right)  \Big] \psi_b
   \nonumber\\
&  - \frac{i}{2}\sum_{b\in \Delta}\Bar{\psi}_{b\mu}\Big[\varepsilon ^{\mu \nu \rho \lambda }\gamma_5\gamma _\nu  \partial_\rho- \gamma^0\left(g_{\omega b}\omega_0 + g_{\rho b} I_{3b} \rho_{03} \right) -\left(m_b-g_{\sigma b}\sigma_0 \right)\varsigma ^{\mu \lambda }\Big]\psi_{b\nu}\nonumber\\
&
+\sum_\lambda\Bar{\psi}_\lambda\left(i\gamma^\mu\partial_\mu-m_\lambda\right)\psi_\lambda- \frac{1}{2} m_\sigma^2 \sigma_0^2  +\frac{1}{2} m_\omega^2 \omega_0^2 +\frac{1}{2} m_\phi^2 \phi_0^2 +\frac{1}{2} m_\rho^2 \rho_{03}^2 \label{lagrangian}
\end{align}
\end{widetext}
where the first sum represents the Dirac-type interacting Lagrangian for the spin-1/2 baryon octet ($H=\{n,p,\Lambda,\Sigma^-,\Sigma^0,\Sigma^+,\Xi^-,\Xi^0\}$) and the second sum represents the Rarita-Schwinger--type interacting Lagrangian for the particles of the spin-3/2 baryon decuplet ($\Delta=\Delta^-,\Delta^0,\Delta^+,\Delta^{++}\}$), where $\varepsilon ^{\mu \nu \rho \lambda}$ is the Levi-Cicita symbol, $\gamma_5=i\gamma _0\gamma _1\gamma _2\gamma _3$ and $\varsigma ^{\mu \lambda }=\frac{i}{2}\left [ \gamma ^\mu,\gamma ^\lambda  \right ]$.
We point to the fact that spin-$3/2$ baryons are described by the Rarita-Schwinger Lagrangian density and that their vector-valued spinor has additional components when
compared to the four components in the spin-$1/2$ Dirac spinors, but, as shown in \cite{DePaoli}, spin-$3/2$ equations of motion can be written compactly as the spin-$1/2$ ones in the RMF regime. 
The last sum describes the leptons admixed in the hadronic matter as a free non-interacting fermion gas ($\lambda=\{e,\mu\}$), as their inclusion is necessary in order to ensure the $\beta$-equilibrium and charge neutrality essential to stellar matter. The remaining terms account for the purely mesonic part of the Lagrangian.

 In DD-RMF models, the coupling constants can be either dependent on the scalar density $n_s$ or the vector density $n_B$, but usually, the vector density parameterizations are considered which influences only the self-energy instead of the total energy \cite{PhysRevLett.68.3408}.
In  this study, we use the DD-RMF parametrization known as DDME2~\cite{ddme2}, where the meson couplings are scaled
with the baryonic density factor $\eta =n_B/n_0$ obeying the function
\begin{equation}
    g_{i b} (n_B) = g_{ib} (n_0) \frac{a_i +b_i (\eta + d_i)^2}{a_i +c_i (\eta + d_i)^2} 
\end{equation}
for $i=\sigma, \omega, \phi$ and 
\begin{equation}
    g_{\rho b} (n_B) = g_{ib} (n_0) \exp\left[ - a_\rho \big( \eta -1 \big) \right],
\end{equation}
for $i=\rho$. The model parameters are fitted from experimental constraints of nuclear matter at or around the saturation density $n_0$, namely the binding energy $B/A$, compressibility modulus $K_0$, symmetry energy $S_0$, and its slope $L_0$, shown in Table~\ref{T1}
~\cite{dutra2014,ddme2}. \\

\begin{center}
\begin{table}[ht]
		\caption{DDME2 parameters (top) and its predictions to the nuclear matter at saturation density (bottom).\label{T1} }
\begin{tabular}{ c c c c c c c }
\hline
$i$ & $m_i(\text{MeV})$ & $a_i$ & $b_i$ & $c_i$ & $d_i$ & $g_{i N} (n_0)$\\
 \hline
 $\sigma$ & 550.1238 & 1.3881 & 1.0943 & 1.7057 & 0.4421 & 10.5396 \\  
 $\omega$ & 783 & 1.3892 & 0.9240 & 1.4620 & 0.4775 & 13.0189  \\
 $\rho$ & 763 & 0.5647 & --- & --- & --- & 7.3672 \\
 \hline
\end{tabular}

\vspace{10pt}

\begin{tabular}{c|cc}
\hline 
Quantity & Constraints \cite{dutra2014, Oertel:2016bki} & This model\\\hline
$n_0$ ($fm^{-3}$) & 0.148--0.170 & 0.152 \\
 $-B/A$ (MeV) & 15.8--16.5  & 16.4  \\ 
$K_0$ (MeV)& 220--260   &  252  \\
 $S_0$ (MeV) & 31.2--35.0 &  32.3  \\
$L_0$ (MeV) & 38--67 & 51\\
\hline
\end{tabular}
\label{T1}
\end{table}
\end{center}

The model-free parameters are fitted considering pure nucleonic (protons and neutrons only) matter. In order to determine the meson couplings to other hadronic species we define the ratio of the baryon coupling to the nucleon one as $\chi_{ib}=g_{i b}/g_{i N}$, with $i = \{\sigma,\omega,\phi,\rho\}$. In this work, we consider hyperons and/or deltas inclusive in the nucleonic matter and
 follow the proposal of ~\cite{Lopes1} to determine their respective $\chi_{ib}$ ratios. It is made through a unified approach relying on symmetry arguments such as the fact that the Yukawa couplings terms present in the Lagrangian density of the DD-RMF models must be invariant under SU(3) and SU(6) group transformations. Hence, the couplings can be fixed to reproduce the potentials  $U_\Lambda =-28$~MeV, $U_\Sigma= 30$~MeV, $U_\Xi=-4$~MeV and $U_\Delta\approx -98$~MeV in terms of a single free parameter $\alpha_v$. Our choice of $\alpha_v=1.0$ for the baryon-meson coupling scheme corresponds to an unbroken SU(6) symmetry, and the values of $\chi_{ib}$ are shown in Table~\ref{T2} taking into account the isospin projections in the lagrangian terms \cite{issifu}.

\begin{center}
\begin{table}[!h]
\caption {Baryon-meson coupling constants  $\chi_{ib}$ \cite{Lopes1}.
\label{T2}}
\begin{tabular}{ c c c c c } 
\hline
 b & $\chi_{\omega b}$ & $\chi_{\sigma b}$ & $I_{3b}\chi_{\rho b}$ & $\chi_{\phi b}$  \\
 \hline
 $\Lambda$ & 2/3 & 0.611 & 0 & 0.471  \\  
  $\Sigma^{-}$,$\Sigma^0$, $\Sigma^{+}$ & 2/3 & 0.467 & $-1$, 0, 1 & -0.471 \\
$\Xi^-$, $\Xi^0$  & 1/3 & 0.284 & $-1/2$, 1/2 & -0.314 \\
  $\Delta^-$, $\Delta^0$   & 1 & 1.053 & $-3/2$, $-1/2$, 1/2, 3/2 & 0  \\
  \hline
\end{tabular}
\end{table}
\end{center}

From the Lagrangian, thermodynamic quantities can be calculated in the standard way for RMF models. The baryonic and scalar densities of a baryon of the species $b$ are given, respectively, by
\begin{equation}
n_b = \frac{\lambda_b}{2\pi ^{2}}\int_{0}^{{k_F}_b}dk\, k^{2}=\frac{\lambda_b}{6\pi ^{2}}{k_F}_b^{3}, \label{eq:rhobarion}
\end{equation} 
and
\begin{equation}
 n^s_b=\frac{\lambda_b}{2\pi ^{2}}\int_{0}^{{k_F}_b} dk \frac{k^{2}m_b^\ast}{\sqrt{k^{2}+{m_b^\ast}^{2}}}, \label{eq:rhoscalar}
\end{equation} 
with ${k_F}$ denoting the Fermi momentum,  since we assume the stellar  matter to be at zero temperature, and $\lambda_b$ is the spin degeneracy factor (2 for the baryon octet and 4 for the deltas).
The effective masses are 
\begin{equation}
    m_b^\ast =m_b- g_{\sigma b} \sigma_0 .
\end{equation}

The energy density is given by
\begin{widetext}
\begin{align}\label{1a}
\varepsilon_B={}& \sum_b \frac{\gamma_b}{2\pi^2}\int_0^{{k_{F}}_b} dk k^2 \sqrt{k^2 + {m_b^\ast}^2}+ \sum_\lambda \frac{1}{\pi^2}\int_0^{{k_{F}}_\lambda} dk k^2 \sqrt{k^2 + m_\lambda^{2}}+ \frac{m_\sigma^2}{2} \sigma_0^2+\frac{m_\omega^2}{2} \omega_0^2 +\frac{m_\phi^2}{2} \phi_0^2  + \frac{m_\rho^2}{2} \rho_{03}^2 .
\end{align}
\end{widetext}
The effective chemical potentials read
\begin{equation}
      \mu_b^\ast = \mu_b- g_{\omega b} \omega_0 - g_{\rho b} I_{3b} \rho_{03} - g_{\phi b} \phi_0 - \Sigma^r,
\end{equation}
where $\Sigma^r$ is the rearrangement term due to the density-dependent couplings
\begin{align}
    \Sigma^r ={}& \sum_b \Bigg[ \frac{\partial g_{\omega b}}{\partial n_b} \omega_0 n_b + \frac{\partial g_{\rho b}}{\partial n_b} \rho_{03} I_{3b}  n_b+ \frac{\partial g_{\phi b}}{\partial n_b} \phi_0 n_b \nonumber \\
    &- \frac{\partial g_{\sigma b}}{\partial n_b} \sigma_0 n_b^s\Bigg],
\end{align}
and the $\mu_b$ are determined by the chemical equilibrium condition
\begin{equation}
    \mu_b=\mu_n-q_b\mu_e, \label{beta}
\end{equation}
in terms of the chemical potential of the neutron and the electron, with $\mu_\mu=\mu_e$. The particle populations of each individual species are determined by Eq.~\eqref{beta} together with the charge neutrality condition $\sum_i n_iq_i=0$, where $q_i$ is the charge of  the baryon or lepton $i$.
The pressure, finally, is given by
\begin{equation}
    P =\sum_i \mu_i n_i - \epsilon + n_B \Sigma^r,
\end{equation}
which receives a correction from the rearrangement term to guarantee thermodynamic consistency and energy-momentum conservation~\cite{Typel1999, PhysRevC.52.3043}.


\subsection{Radial oscillations}
\label{radial}

Einstein's equations of General Relativity govern the structure and dynamical evolution of NSs because of their intense gravitational field. Moreover, the static equilibrium structure-based Einstein field equation can be used to calculate the radial oscillation properties \cite{1966ApJ...145..505B}. Consider a spherically symmetric system with only radial motion, where the metric is now time-dependent. For radial displacement $\Delta r$ with $\Delta P$ as the perturbation of the pressure, the small perturbation 
 of the equations governing the dimensionless quantities $\xi$ = $\Delta r/r$ and $\eta$ = $\Delta P/P$ are defined as \cite{1977ApJ...217..799C, 1997A&A...325..217G}
 \begin{equation}\label{ksi}
     \xi'(r) = -\frac{1}{r} \Biggl( 3\xi +\frac{\eta}{\gamma}\Biggr) -\frac{P'(r)}{P+\mathcal{E}} \xi(r),
 \end{equation}
 \begin{equation}\label{eta}
 \begin{split}
          \eta'(r) = \xi \Biggl[ \omega^{2} r (1+\mathcal{E}/P) e^{\lambda - \nu } -\frac{4P'(r)}{P} -8\pi (P+\mathcal{E}) re^{\lambda} \\
     +  \frac{r(P'(r))^{2}}{P(P+\mathcal{E})}\Biggr] + \eta \Biggl[ -\frac{\mathcal{E}P'(r)}{P(P+\mathcal{E})} -4\pi (P+\mathcal{E}) re^{\lambda}\Biggr] ,
      \end{split}
 \end{equation}
 where $\omega$ is the frequency oscillation mode and $\gamma$ is the adiabatic relativistic index defined as
 \begin{equation}
     \gamma = \Biggl( 1+\frac{\mathcal{E}}{P}\Biggr) c_s^{2} ,
 \end{equation}
 where $c_s^{2}$ is the speed of sound squared
  \begin{equation}\label{cs}
     c_s^{2} = \Biggl(\frac{dP}{d\mathcal{E}}\Biggr)c^{2} .
 \end{equation}

The two coupled differential equations Eqs.~\ref{ksi} and \ref{eta} are supplemented with two additional boundary conditions, one at the center where $r$ = 0, and another at the surface where $r$ = $R$. The boundary condition at the center requires that
\begin{equation}
    \eta = -3\gamma \xi 
\end{equation}
must be satisfied. The equation Eq.~\eqref{eta} must be finite at the surface and hence
\begin{equation}
    \eta = \xi \Biggl[ -4 +(1-2M/R)^{-1} \Biggl( -\frac{M}{R} -\frac{\omega^{2} R^{3}}{M}\Biggr)\Biggr]
 \end{equation}
 must be satisfied where $M$ and $R$ correspond to the mass and radius of the star, respectively. The frequencies are computed by
 \begin{equation}
\nu = \frac{\bar{\omega}}{2\pi}~~(kHz),
\end{equation}
where $\bar{\omega}$ = $\omega t_0$ is the dimensionless quantity computed at $t_0$ = 1 ms.  

 These equations represent the Sturm-Liouville eigenvalue equations for $\omega$. The solutions provide the discrete eigenvalues $\omega_n ^{2}$ and can be ordered as 
 \begin{equation*}
\omega_0 ^{2} < \omega_1 ^{2} <... <\omega_n ^{2}, 
 \end{equation*}
 where $n$ is the number of nodes for a given NS. For a real value of $\omega$, the star will be stable and for an imaginary frequency, it will become unstable. Also, since the eigenvalues are arranged in above defined manner, it is important to know the fundamental $f$-mode frequency ($n$ = 0) to determine the stability of the star.  

\section{Numerical results and discussion}
\label{results}

\subsection{EoS and MR Profile}
\label{mr}
\begin{figure}[ht]
	\includegraphics[scale=0.35]{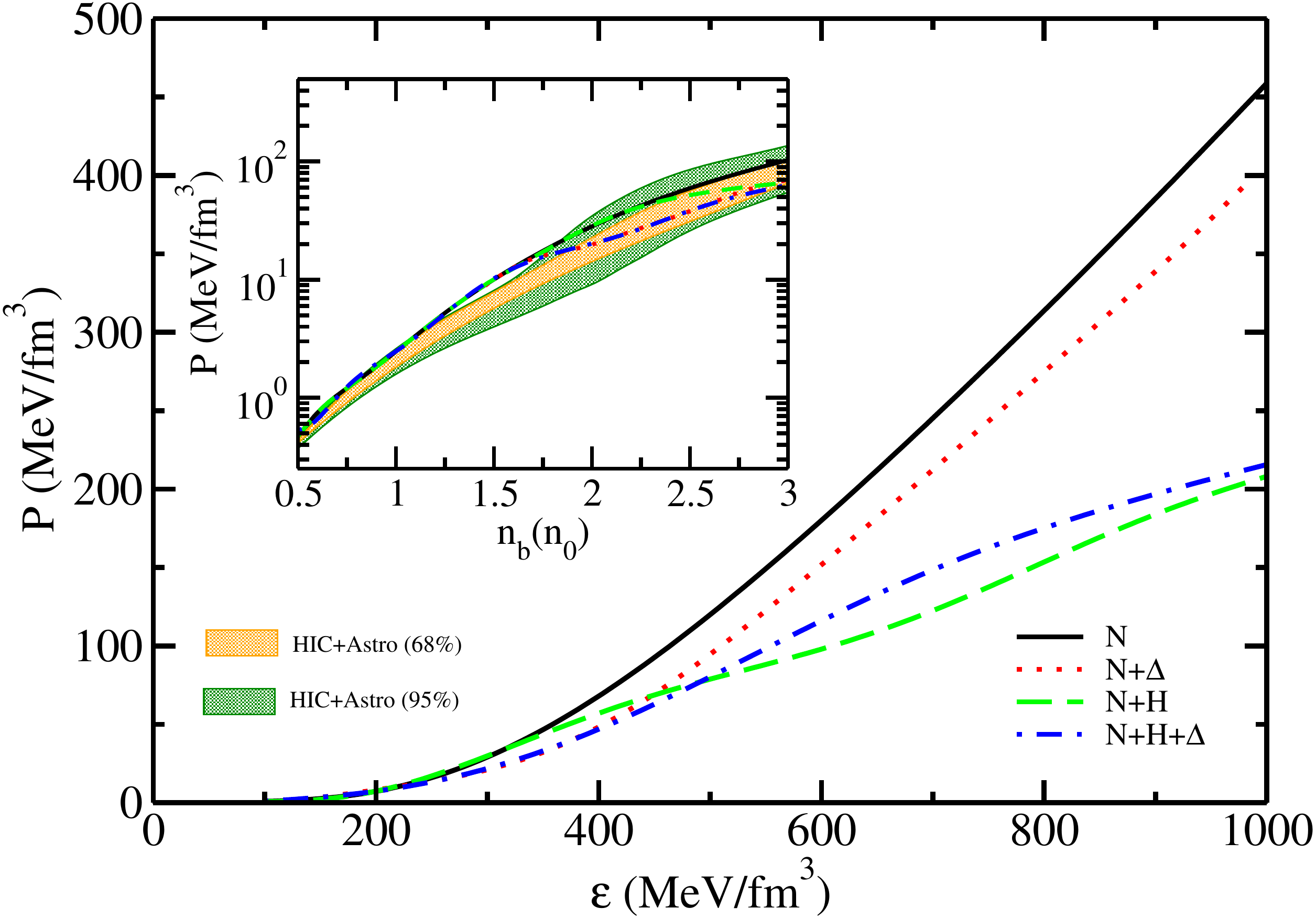}
	\caption{(color online) Energy density and pressure variation for the given DD-ME2 parameter set. The solid line represents the pure nucleonic matter (N) while dotted, dashed, and dash-dotted lines represent the EoS for $\Delta$-inclusive nuclear matter (N+$\Delta$), hyperons (N+H), and $\Delta$-inclusive hyperonic matter (N+H+$\Delta$), respectively, for $\alpha_v$ = 1.0. The inset plot shows the number density vs pressure variation for different matter compositions. The orange (68\%) and green (95\%) shaded regions show the joint constraints from the heavy-ion collision (HIC) experiments and multi-messenger astrophysics (Astro) \cite{Huth2022}. }
	\label{fig1} 
\end{figure}

Fig.~\ref{fig1} shows the variation of pressure with energy density (EoS) for an NS in beta-equilibrium and charge-neutral conditions. The pure nucleonic matter produces stiff EoS in the high-density region. The EoS softens when $\Delta$s are added to the nuclear matter. This is because the presence of more degrees of freedom distributes the Fermi pressure among the many particles as a result of the inclusion of new particles, softening the EoS. However, we must point out the fact that only $\Delta^{0}$ and $\Delta^{-}$ baryons are considered in the $\Delta$-inclusive nuclear and $\Delta$-inclusive hyperonic matter because the inclusion of $\Delta^{+}$ and $\Delta^{++}$ baryons allows the nucleon effective mass to drop to zero for very low densities and hence precluding the neutron stars from achieving densities high enough to describe the maximum mass star. A more detailed explanation of such behavior of $\Delta$ baryons is explained in Ref.\cite{PhysRevC.106.055801}, where the authors show that the increase of the exotic particle abundance adds to the negatively contributing term of the effective nucleon mass, through the scalar density dependence of the $\sigma$ field. This issue was already known for some hyperon matter models, but the fact that the SU(6) coupling scheme enhances very strongly the abundance of resonances makes this behavior very sensible when $\Delta$ baryons are present. As $\Delta^{+}$ and $\Delta^{++}$ baryons ought to be unfavored in the low and intermediate densities due to the charge neutrality condition, excluding them altogether is a possible workaround to that problem. While the hyperons further soften the EoS, the addition of $\Delta$s in the hyperonic matter, N+H+$\Delta$, is more complex. As seen from Fig.~\ref{fig1}, at lower densities, the N+H+$\Delta$ is softer than the N+H composition. With the increase in the density, the EoS with the N+H+$\Delta$ composition becomes stiffer than the N+H composition. The explanation for this is that the appearance of $\Delta^{-}$ baryon replaces a neutron-electron pair at the top of their Fermi seas which are favored over the light baryons because of the attractive potential. The electric charge-neutral particles, $\Lambda^{0}$ and $\Delta^{0}$, appear later. The inset in Fig.~\ref{fig1} shows the number density vs pressure variation for different matter compositions. The joint constraints from the heavy-ion collision (HIC) experiments and multi-messenger astrophysics (Astro), orange (68\%), and green (95\%) credible ranges are also shown \cite{Huth2022}. As we can see, the EoSs nearly satisfy the joint constraints at a lower density. The appearance of delta baryons ensures that all the EoSs satisfy these constraints. For the unified EoS, the Baym-Pethick-Sutherland (BPS) EoS \cite{Baym:1971pw} is used for the outer crust part. For the inner crust, the EoS in the non-uniform matter is generated by using the DD-ME2 parameter set in Thomas-Fermi approximation \cite{PhysRevC.79.035804, PhysRevC.94.015808,rather2020effect}.  \\

\begin{figure}[h]
  	\includegraphics[scale=0.35]{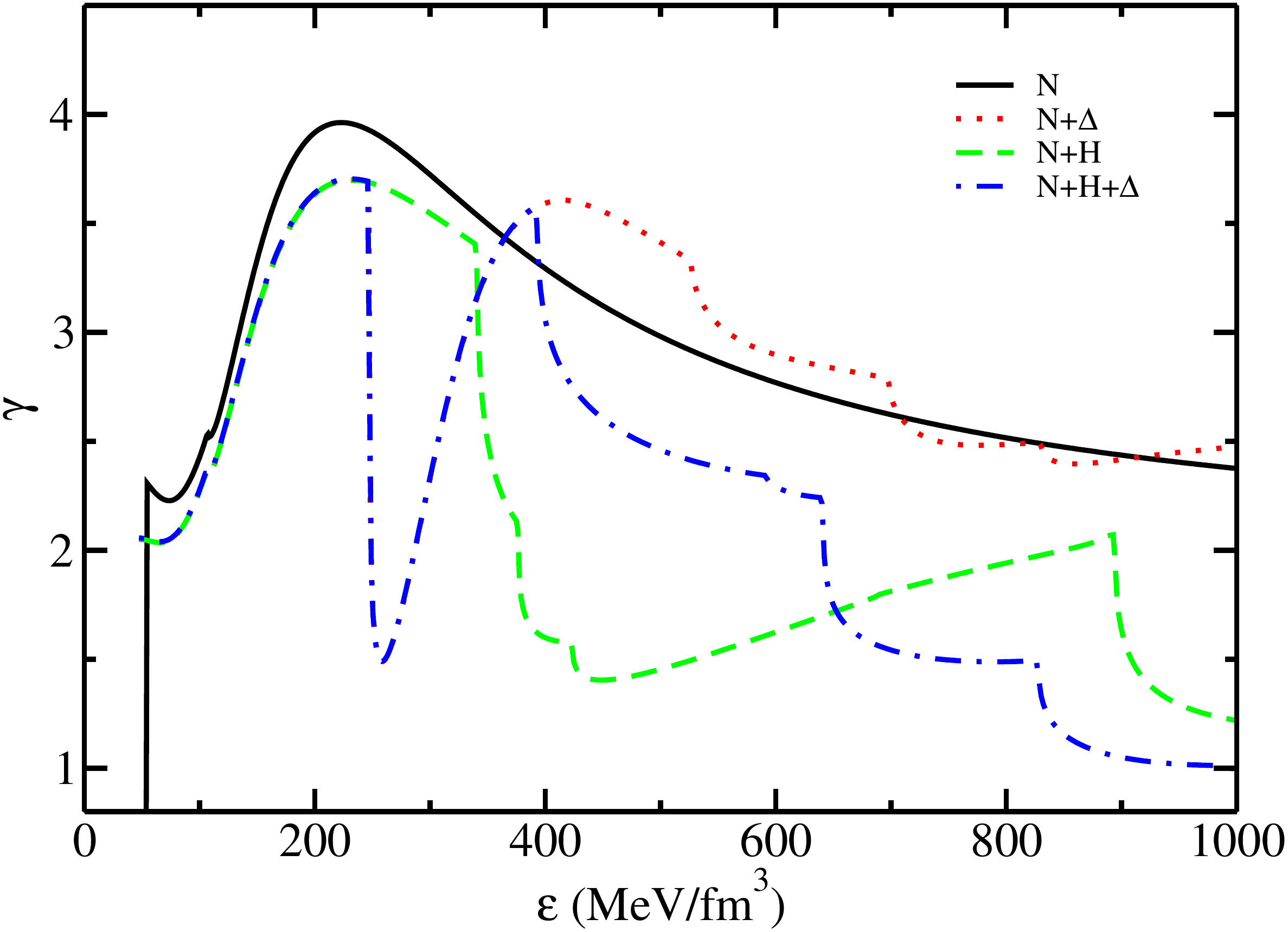}
  	\caption{(Color online) Adiabatic index as a function of energy density for the DD-ME2 parameter set with the pure nuclear matter,  $\Delta$-inclusive nuclear matter, hyperonic matter, and  $\Delta$-inclusive hyperonic matter.}
  	\label{fig2a} 
  \end{figure}

Fig.~\ref{fig2a} displays the adiabatic index $\gamma$ vs. the energy density for different matter compositions. For pure nucleonic matter, the $\gamma$ curve increases to a peak value at low energy density and then drops smoothly. The presence of hyperons, especially $\Lambda^0$, softens the EoS and the value of $\gamma$ drops at around $\approx$ 350 MeV/fm$^3$, and each following peak can be associated with the onset of a new particle species. For nuclear and hyperonic matter with $\Delta$ baryons, the value of $\gamma$ drops sharply at around $\approx$ 250 MeV/fm$^3$ due to the onset of $\Delta^-$ baryons. But as the density increases, the $\gamma$ also increases and becomes larger than the pure nucleonic matter. This large increase in the behavior of $\gamma$ is not seen in the hyperonic matter. For $\Delta$-inclusive hyperonic matter, we see a huge drop in the value $\gamma$ due to the $\Delta^-$ threshold followed by a quick increase and then a new drop due to the onset of $\Lambda^0$ hyperon.
  
  \begin{figure}[h]
  	\includegraphics[scale=0.35]{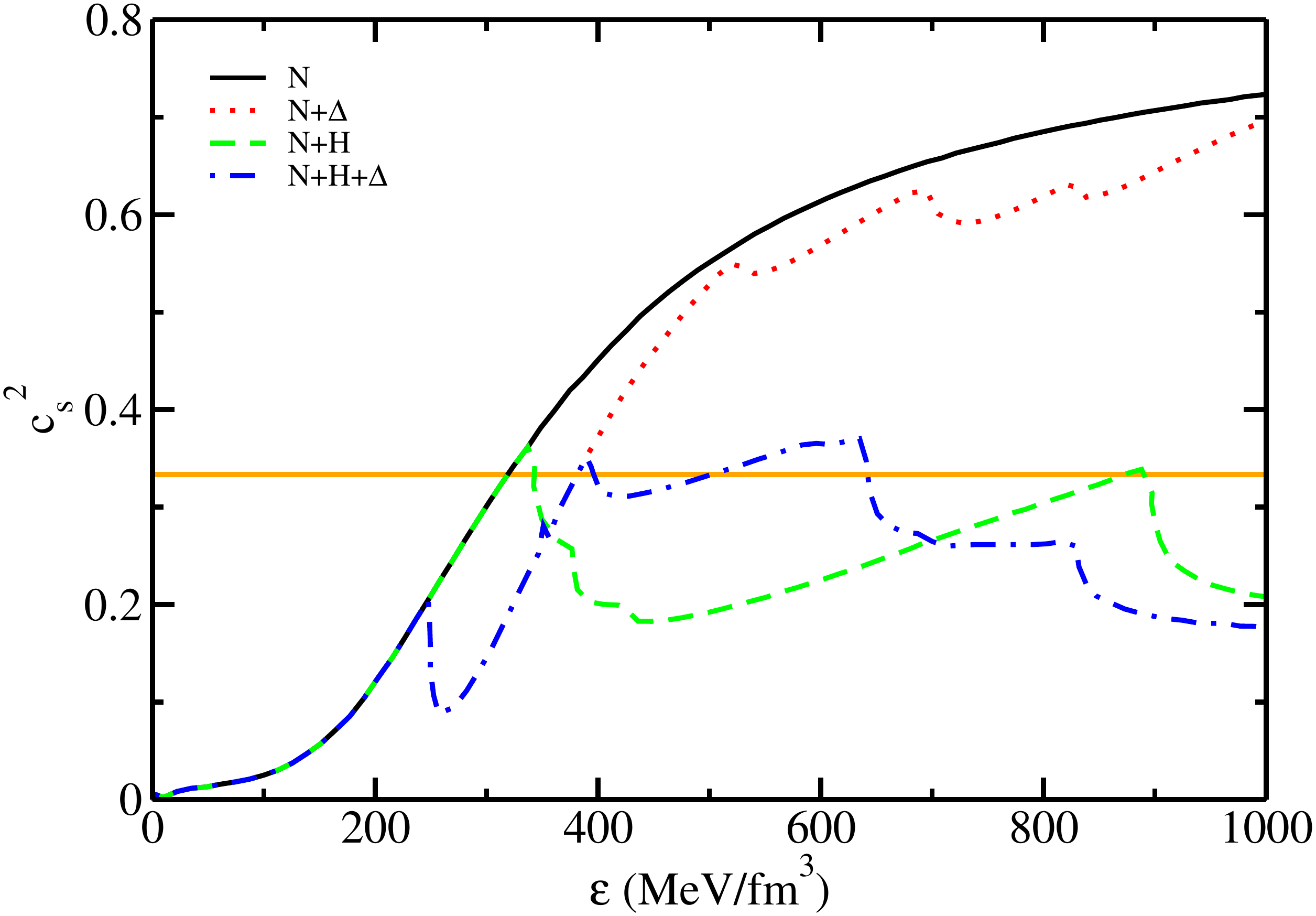}
  	\caption{(Color online) Speed of sound squared as a function of energy density for the DD-ME2 parameter set with the pure nuclear matter,  $\Delta$-inclusive nuclear matter, hyperonic matter, and  $\Delta$-inclusive hyperonic matter. The solid orange line represents the conformal limit $c_s^{2}$ = 1/3.}
  	\label{fig2} 
  \end{figure}
  
  Fig.~\ref{fig2} depicts the behavior of the speed of sound squared as a function of energy density for different compositions of the matter studied in this work. The speed of sound is an important quantity that conveys information about shear viscosity, tidal deformability, and gravitational wave signatures \cite{PhysRevC.102.055801, Lopes_2021}. It is defined as the derivative of pressure with respect to energy density with its square defined by Eq.~\eqref{cs}. It can also be interpreted as a measurement of the stiffness of the EoS, with a higher speed yielding a higher pressure at a given energy density and allowing a larger star mass for a given radius. 
 Thermodynamic stability ensures that $c_s^{2}$ $>$ 0 and causality implies an absolute bound $c_s^{2}$ $\leq$ 1. For very high densities, perturbative QCD findings anticipate an upper limit of $c_s^{2}$ = 1/3 \cite{PhysRevLett.114.031103}. The two solar mass requirements, according to several studies \cite{PhysRevLett.114.031103, PhysRevC.95.045801, Tews_2018}, necessitates a speed of sound squared that exceeds the conformal limit ($c_s^{2}$ = 1/3), revealing that the matter inside of NS is a highly interacting system.
  From Fig.~\ref{fig2}, we can see a very large value of the $c_s^{2}$ for the pure nucleonic matter. When different particle compositions are considered, one can see the kinks corresponding to the onset of a new particle species at the same point as the ones in the adiabatic index curves. The conformal limit is violated in the case of pure nucleonic and $\Delta$-inclusive nuclear matter. Also, the curve for N+H+$\Delta$ composition predicts a higher value of the speed of sound squared at intermediate densities because of the early appearance of $\Delta^{-}$ particles, as explained earlier.

  \begin{figure}
 	\includegraphics[scale=0.35]{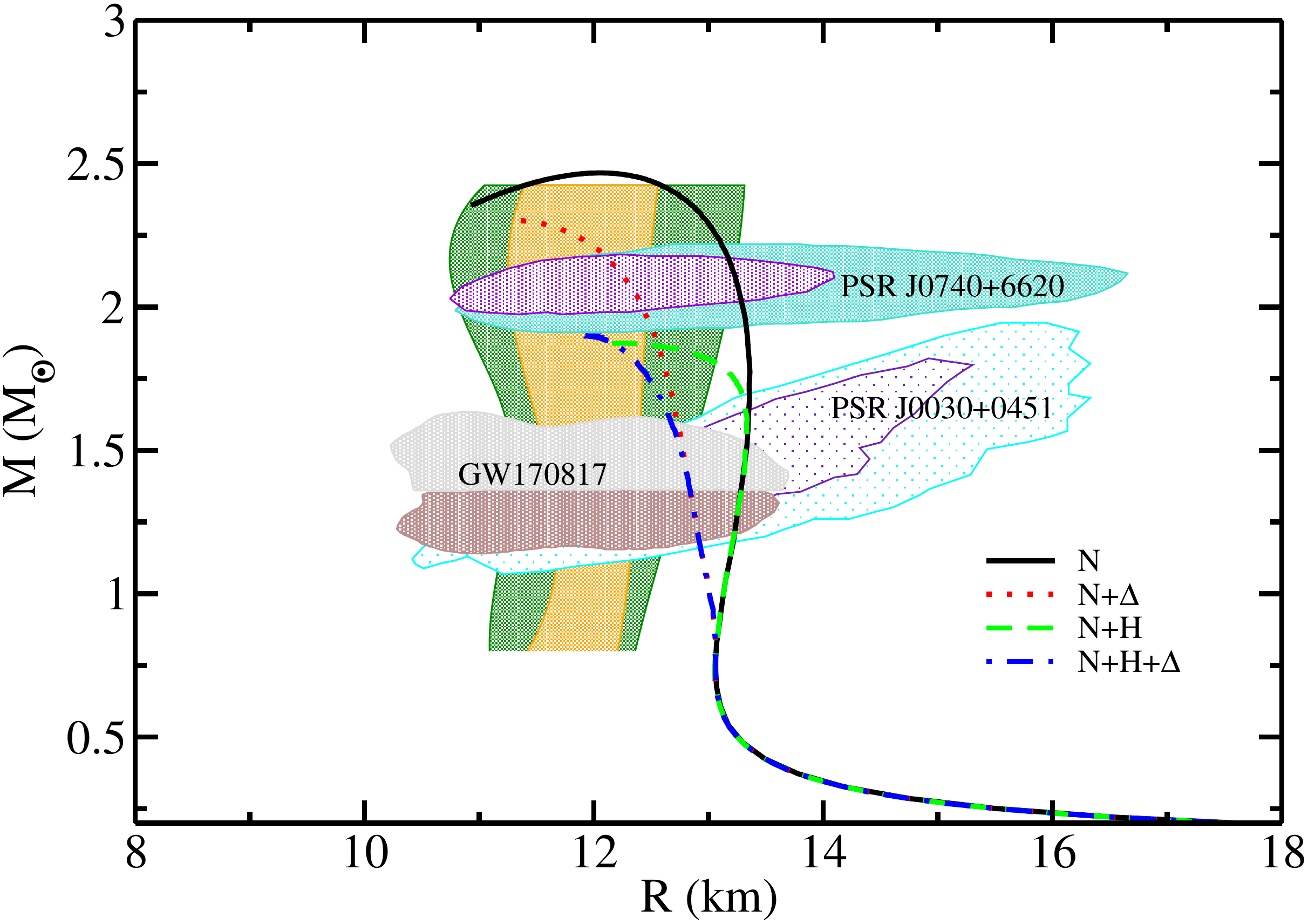}
 	\caption{(Color online) Mass-Radius profile for DD-ME2 parameter set with different compositions of $\Delta$ baryons and hyperons. The solid (dashed) lines represent the MR plot for the pure nucleonic matter. The dotted, dashed, and dash-dotted lines represent the MR profile for $\Delta$-inclusive nuclear matter, hyperons, and $\Delta$-inclusive hyperonic matter, respectively. The 68\% (violet) and 95\% (turquoise) credible regions for mass and radius are inferred from the analysis of PSR J0740+6620 \cite{miller2021,2021ApJ...918L..27R}. For PSR J0030+0451, the indigo dotted region is for 68\% credibility while the cyan dotted region is for 95\% credibility \cite{Miller_2019a}.   The grey upper (brown lower) shaded region corresponds to the higher (smaller) component of the GW170817 event \cite{Abbott_2020a}. The joint constraints from HIC experiments and multi-messenger astrophysics with 68\% (orange) and 95\% (green) credible ranges are also shown \cite{Huth2022}.}
 	\label{fig3} 
 \end{figure}

With the solutions from the TOV equations for different EoSs, Fig.~\ref{fig3} displays the mass-radius profile for different compositions of the matter. For pure nucleonic matter, a maximum mass of 2.46 $M_{\odot}$ is achieved at a radius of 12.05 km. With the $\Delta$ baryons present in the star, the maximum mass and the corresponding radius decrease to a value of 2.24 $M_{\odot}$ and 11.87 km, respectively. The decrease in the mass and the radius of $\Delta$-inclusive nuclear matter depends upon the value of $\alpha_v$. The higher the value of $\alpha_v$, the lower the maximum mass, corresponding radius, and the radius at the canonical mass. In our case with $\alpha_v$ = 1.0, the radius at the canonical mass decrease from 13.29 km for the pure nucleonic matter to 12.82 km for $\Delta$-inclusive nuclear matter. The presence of hyperons softens the EoS and hence the maximum mass decrease to a value of 1.87 $M_{\odot}$ with the corresponding radius of 12.09 km. For the $\Delta$-inclusive hyperonic matter at $\alpha_v$ =1.0, The stiffness of the EoS predicts a maximum mass of 1.90 $M_{\odot}$ with a radius of 11.90 km. The constraints on the mass and the radius from various measurements \cite{miller2021,2021ApJ...918L..27R, Miller_2019a, Abbott_2020a} are very well satisfied by the N and N+$\Delta$ profile, while the N+H and N+H+$\Delta$ profiles nearly satisfy the description of the PSR J0740+6620 for mass and radius  \cite{miller2021,2021ApJ...918L..27R}. The joint constraints from heavy-ion collision (HIC) experiments and multi-messenger astrophysics are very well satisfied by all the mass-radius profiles. Since the maximum mass for N+H and N+H+$\Delta$ EoSs is lower than the mass of the super heavy pulsar PSR J0740+6620, these EoSs can be ruled out. However, for the comparison, we keep the results and calculate the radial profiles with N+H and N+H+$\Delta$ EoSs also to see the effect of delta baryons on the hyperonic matter. 

\subsection{Radial Profiles}
\label{profile}
\begin{figure*}[htbp!]
	\includegraphics[width=0.75\textwidth]{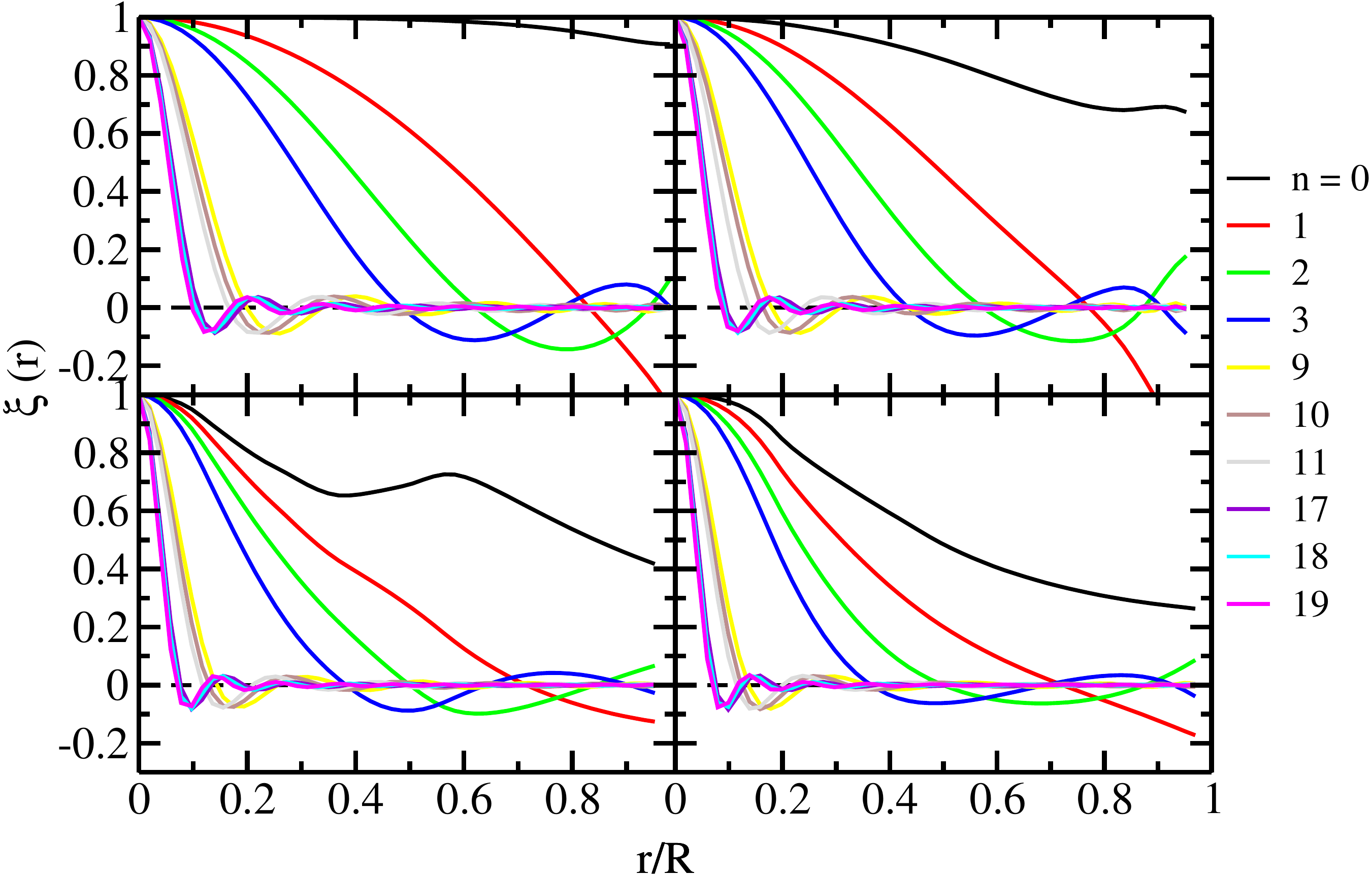}
	\caption{(Color online) The radial displacement perturbation $\xi(r)$ = $\Delta r/r$ as a function of  dimensionless radius distance $r/R$ for lower $f$-mode (n = 0), lower order $p$-modes (n = 1, 2, 3), intermediate $p$-modes (n = 9, 10, 11), and high excited modes (n = 17, 18, 19). The upper left (right) panel represents the result for pure nucleonic ($\Delta$-inclusive nucleonic) matter, while the lower left (right) panel represents the result for hyperonic ($\Delta$-inclusive hyperonic) matter. }
	\label{fig5}
\end{figure*}

\begin{figure*}[htb!]
	\includegraphics[width=0.75\textwidth]{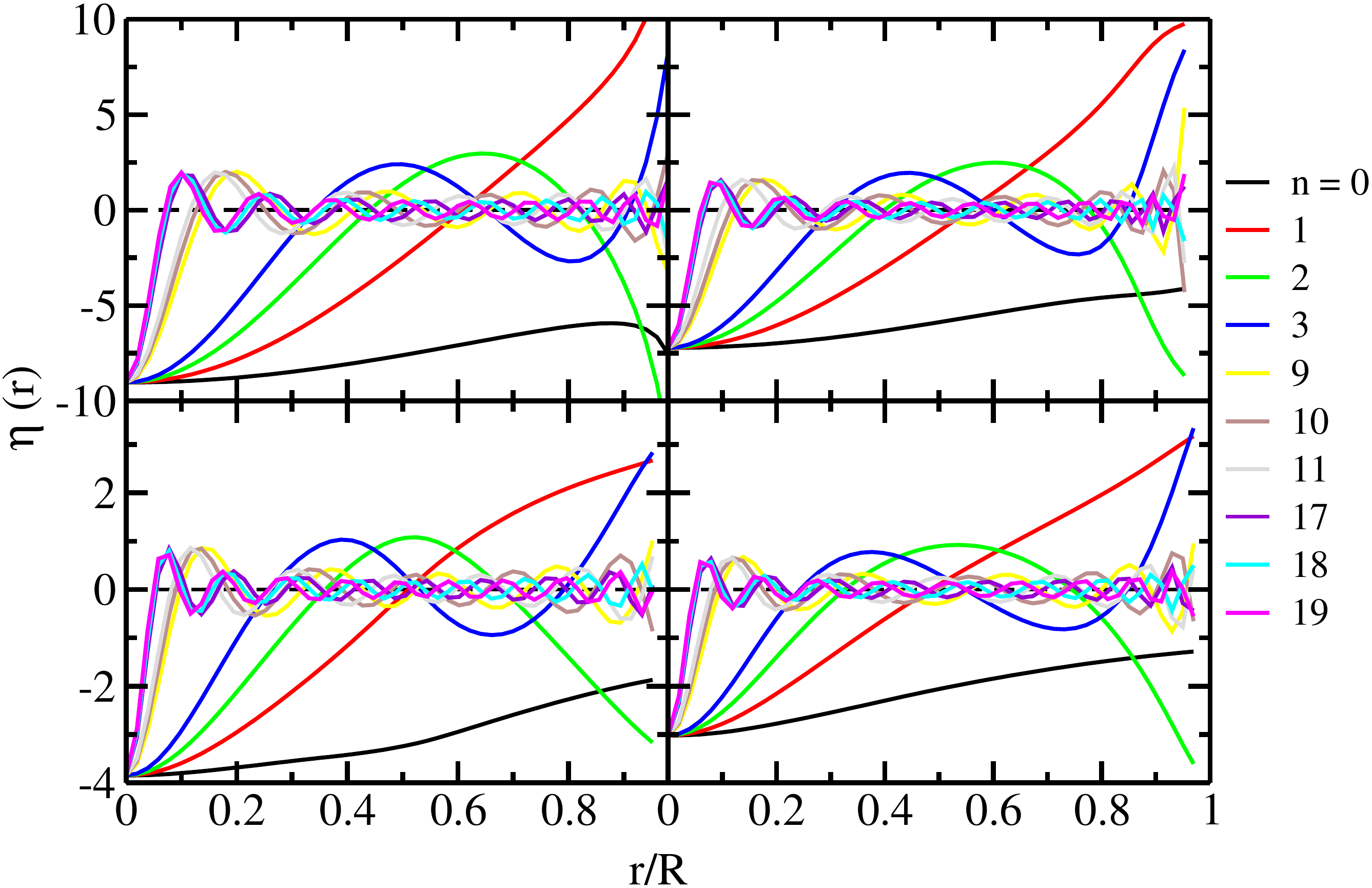}
	\caption{(Color online) The radial pressure perturbation $\eta(r)$ = $\Delta r/r$ as a function of dimensionless radius distance $r/R$ for lower $f$-mode (n = 0), lower order $p$-modes (n = 1, 2, 3), intermediate $p$-modes (n = 9, 10, 11), and high excited modes (n = 17, 18, 19). The upper left (right) panel represents the result for pure nucleonic ($\Delta$-inclusive nucleonic) matter, while the lower left (right) panel represents the result for hyperonic ($\Delta$-inclusive hyperonic) matter.}
	\label{fig6}
\end{figure*}

The radial displacement perturbation profile $\xi(r)$ and pressure perturbation profile $\eta(r)$ as a function of  dimensionless radius distance $r/R$ is plotted in Figs.~\ref{fig5} and \ref{fig6}, respectively. These profiles are plotted for different particle compositions, pure nucleonic (upper left), $\Delta$-inclusive nucleonic (upper right), hyperonic (lower left), and $\Delta$-inclusive hyperonic (lower right) matter at the corresponding maximum masses (with different central densities). Only the $f$-mode (n = 0), lower order $p$-modes (n = 1, 2, 3), intermediate (n = 9, 10, 11), and high excited modes (n = 17, 18, 19) are shown. In the region $0<r<R$, exactly $n$ nodes are obtained for the $n$th mode both for $\xi$ and $\eta$ profiles, thereby following the Sturm-Liouville system. From Fig.~\ref{fig5}, one can see that the amplitude of $\xi_n(r)$ for each frequency mode $\nu_n$ is larger near the center and small at the surface. The lower modes show a smooth drop in their profiles while the higher modes depict small oscillations which would become large for higher modes. For $\Delta$-inclusive nucleonic matter, we see a small kink at around $r/R$ = 0.8 for $n$ = 0 mode. For hyperonic matter, the kink at the same node is large and present at around $r/R$ = 0.3. These kinks in $\xi(r)$ represent the emergence of new exotic particles which provides a discontinuity in the adiabatic index, that appears in Eq.~\eqref{ksi} explicitly. One can see a rapid sign change near the center of the star which along with the amplitude decrease as one moves toward the surface of the star. From Fig.~\ref{fig6}, it's observed that the amplitude of $\eta_n(r)$ is larger near the center and also at the surface of the star. Although the $\eta$ oscillations are directly proportional to the Lagrangian pressure variation $\Delta P$, the amplitude of $\eta_n(r)$ for consecutive $n$ have large amplitudes near the surface, and hence the contribution from $\eta_{n+1}$ - $\eta_n$ cancels out because of the opposite signs, thereby satisfying the condition that $P(r = R)$ = 0. This implies that $\eta_{n+1}$ - $\eta_n$  and also $\xi_{n+1}$ - $\xi_n$ are more sensitive to the star's core.  As a result, the measurement of $\Delta \nu_n$ = $\nu_{n+1}$ - $\nu_n$  is an observational imprint of this star's innermost layers.

\begin{center}
\begin{table}[ht]
		\caption{20 lowest order radial oscillation frequencies, $\nu$ in (kHz) for different EoSs considered. For each EoS, the frequencies are calculated at the maximum mass of the corresponding star.\label{table1} }
\begin{tabular}{ p{1.5cm}p{1.5cm}p{1.5cm}p{1.5cm}p{1.5cm} }
 \hline
\multirow{2}{*}{Nodes} &\multicolumn{4}{c}{EoS} \\
 \cline{2-5}
  & N&N+$\Delta$ & N+H & N+H+$\Delta$\\
 \hline
 0&0.571&	1.578&	1.977	&1.865\\
1	&5.173	&5.891	&6.582	&6.241\\
2	&8.163	&9.118	&10.497	&9.656\\
3	&11.060	&12.379	&14.231	&13.390\\
4	&13.960	&15.674	&17.907	&16.652\\
5	&16.878	&18.984	&21.527	&19.986\\
6	&19.812	&22.316	&25.064	&23.422\\
7	&22.758	&25.649	&28.758	&27.116\\
8	&25.713	&29.011	&32.317	&30.754\\
9	&28.674	&32.358	&35.633	&34.492\\
10	&31.639	&35.732	&39.177	&38.135\\
11	&34.609	&39.089	&42.800	&41.779\\
12	&37.580	&42.457	&46.347	&45.505\\
13	&40.556	&45.822	&50.016	&49.174\\
14	&43.535	&49.189	&53.464	&52.622\\
15	&46.518	&52.556	&57.117	&56.275\\
16	&49.500	&55.933	&60.527	&59.685\\
17	&52.483	&59.308	&64.180	&63.380\\
18	&55.469	&62.688	&67.796	&67.054\\
19	&58.457&	66.061	&71.603	&70.618\\
 \hline
\end{tabular}
\end{table}
\end{center}

\begin{figure*}[htbp!]
	\includegraphics[scale=0.50]{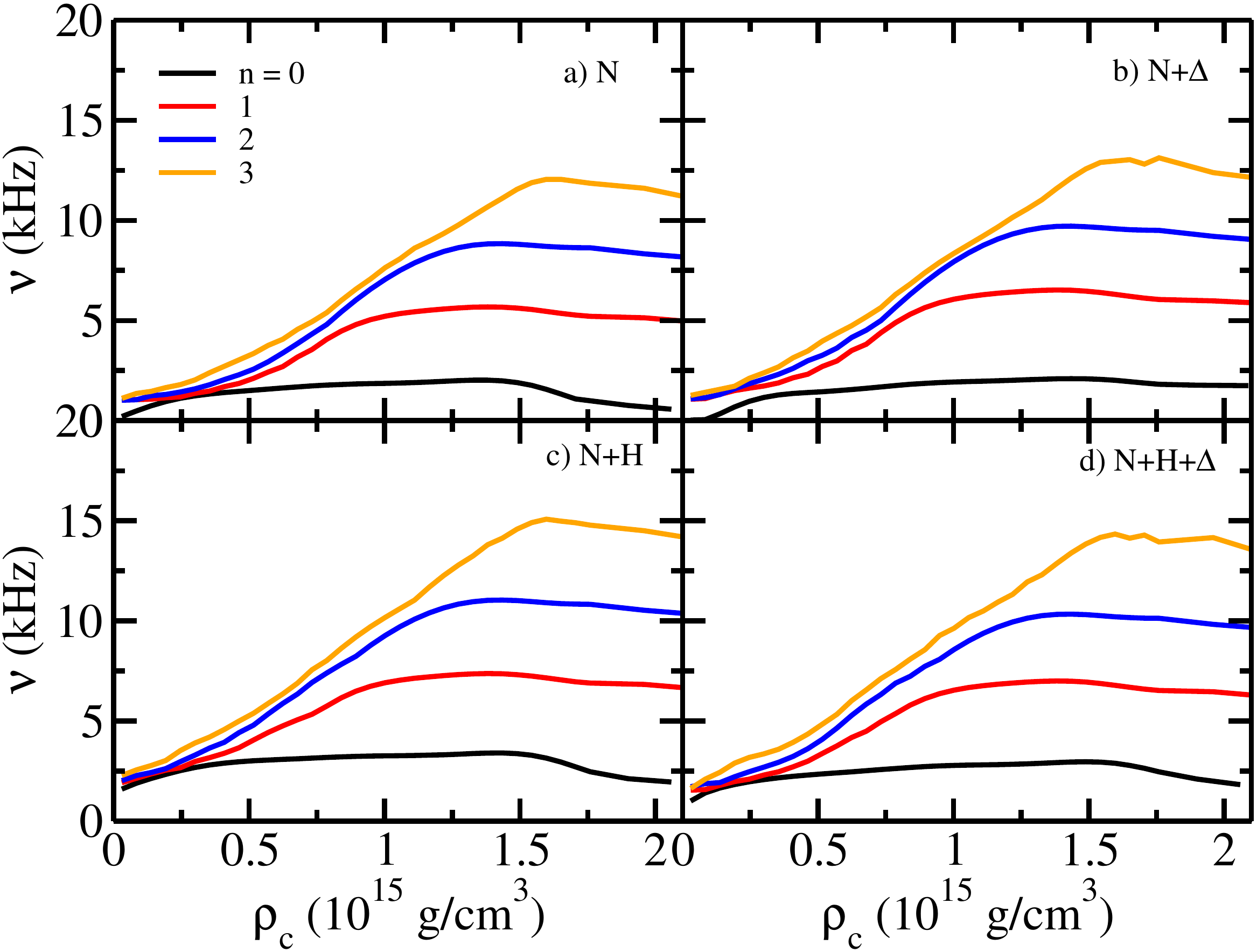}
	\caption{(Color online) Frequencies of radially oscillating NS as a function of central energy density for a) pure nucleonic matter (N), b) $\Delta$-inclusive nucleonic matter (N+$\Delta$), c) hyperonic matter (N+H), and d) $\Delta$-inclusive hyperonic matter (N+H+$\Delta$). The frequencies for lower radial modes ($n$ = 0, 1, 2, 3) are shown.}
	\label{fig7} 
\end{figure*}

Table \ref{table1} displays the frequencies, $\nu$ in kHz, of the first 20 nodes for pure nucleonic matter, $\Delta$-inclusive nucleonic matter, hyperonic matter, and $\Delta$-inclusive hyperonic matter, respectively. All these frequencies are obtained at the corresponding maximum masses of the EoSs. The node $n$ = 0 corresponds to the $f$-mode frequency while the others correspond to the lower and highly excited $p$-modes. The frequency of the $f$-mode for pure nucleonic EoS is lower as compared to the other EoSs with deltas and hyperons.

Fig.~\ref{fig7} shows the frequencies of radially oscillating NS with different matter compositions, as a function of central energy density for lower radial modes, $n$ = 0, 1, 2, and 3. It is clear from the figure that for the same core density, stellar models with softer EoSs exhibit higher $f$-mode frequencies than those of stiffer EoSs. The stellar models of softer EoSs are typically linked to larger average densities and more centrally compressed stars. The star is getting closer to its stability limit as the center density rises and the $f$-mode frequency ($n$ = 0) begins to shift toward zero at the same moment. An eigenmode with zero frequency is a characteristic of the stability limit itself. The $f$-mode frequency of N+$\Delta$ and N+H EoSs is higher as compared to the pure nucleonic matter because of the delta and hyperonic that make the EoS softer. Since the N+H+$\Delta$ is stiffer than the N+H EoS, the corresponding $f$-mode frequency is lower.

Higher modes oscillate more frequently than lower stable modes do, and for all modes, this frequency appears to decrease as the center energy density approaches the minimum value of the specific star model. The explanation for this comes from the fact that the NSs at very high densities can be approximated as being homogeneous and thus the angular frequency $\omega_0^{2}$ follows the relation $\omega_0^{2}$= $\rho (4-3\gamma)$ \cite{1977ApJS...33..415A,10.1093/mnras/stac2622}. From Fig.~\ref{fig7}, it is observed that with $\Delta$s in the pure nucleonic and hyperonic matter, the higher modes show small kinks. This illustrates an essential observation that leads to a series of ``avoided crossings'' between the various modes: the frequencies of two subsequent modes from different families reject each other as they approach one another \cite{1997A&A...325..217G,kokkostas}. This ``avoided crossing'' is a characteristic of a realistic EoS \cite{kokkostas} and is present in all four cases at lower densities. 

\begin{figure}[ht]
	\includegraphics[scale=0.35]{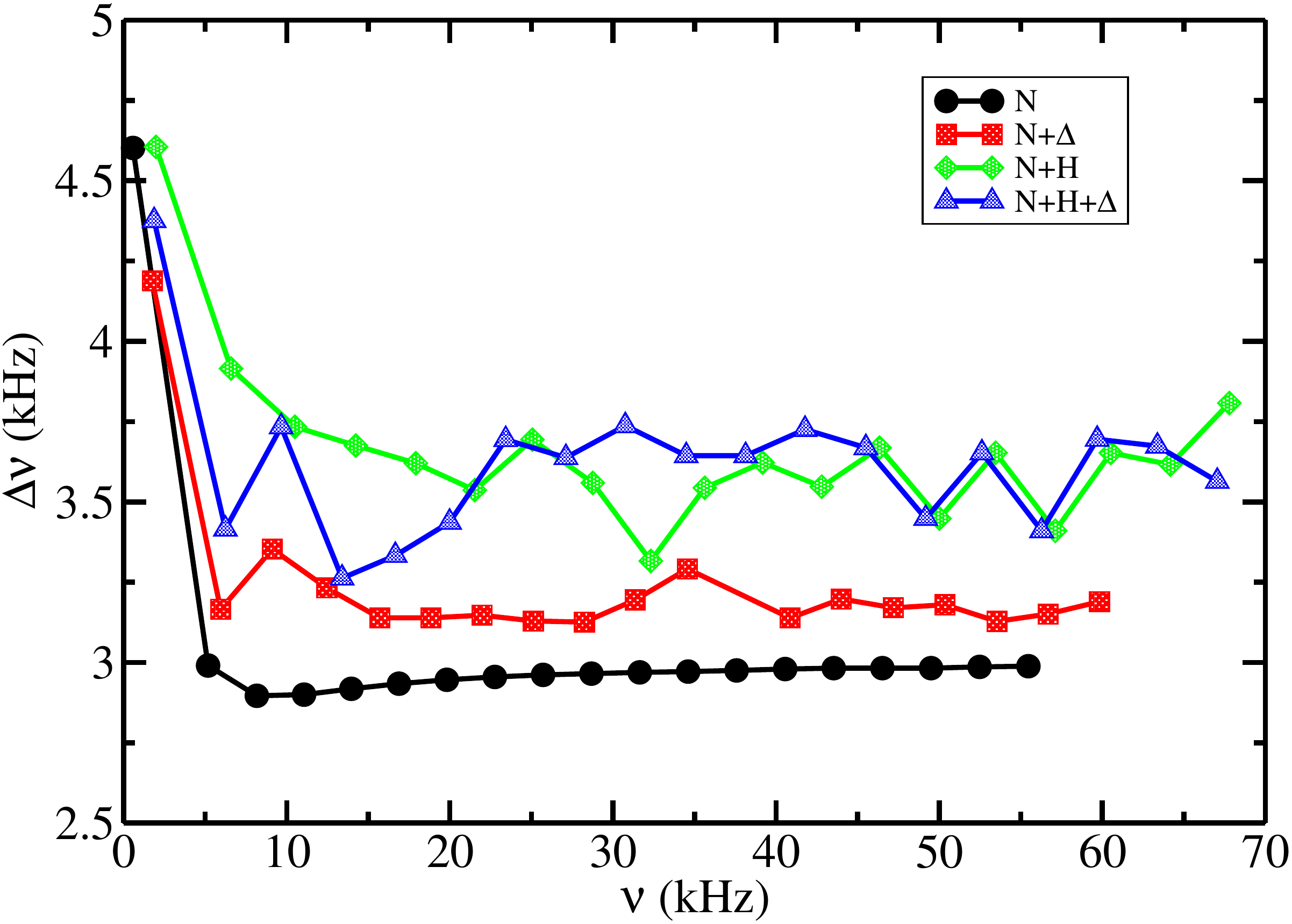}
	\caption{(Color online) Frequency difference $\Delta \nu_n$ = $\nu_{n+1}$ - $\nu_n$ vs $\nu_n$ in kHz for pure nucleonic (N), $\Delta$-inclusive nucleonic (N+$\Delta$), hyperonic (N+H), and $\Delta$-inclusive hyperonic matter (N+H+$\Delta$) matter. }
	\label{fig8} 
\end{figure}

Fig.~\ref{fig8} displays the frequency difference $\Delta \nu_n$ = $\nu_{n+1}$ - $\nu_n$ vs $\nu_n$ in kHz for pure nucleonic (N), $\Delta$-inclusive nucleonic (N+$\Delta$), hyperonic (N+H), and $\Delta$-inclusive hyperonic matter (N+H+$\Delta$) matter. For pure nucleonic EoS, the separation between the modes is almost the same and there are no fluctuations at lower modes ($n$ = 0). 
While the frequencies in the $\Delta$-inclusive nucleonic matter are higher than pure nucleonic one, as one would expect because of the soft N$\Delta$ EoS than N, the difference between consecutive modes is also the same with minor fluctuations. With the hyperonic and $\Delta$-inclusive hyperonic matter, they oscillate with higher frequencies, and the magnitude of $\Delta \nu_n$ is higher. This shows that the decrease in the central baryon density of the star, and, hence, of its mass leads to a large separation $\Delta \nu_n$. We also observe the erratic fluctuations present in $\Delta \nu_n$ for N+H and N+H+$\Delta$ cases. These fluctuations arise from the significant variation of the speed of sound squared $c_s^{2}$ or the relativistic adiabatic index $\gamma$ on the transition layer separating the inner and outer core of the NS, which has an amplitude proportionate to the magnitude of the discontinuity.
This is also due to the fact that we have considered a unified EoS in the present study.
Although the radial oscillation for the lowest order mode ($n$ = 0) is not highly impacted by the crust because it typically accounts for less than 10\% of the stellar radius and the oscillation nodes are situated far inside the NS core. But other high oscillation modes are present in the crust of the star and hence the eigenfrequencies are modified (characterized by the peaks in the $\Delta \nu_n$) \cite{https://doi.org/10.48550/arxiv.2205.02076}.  
For a given EoS without crust, the variation in the frequency, $\Delta \nu_n$, is smooth as discussed in Refs. \cite{PhysRevD.98.083001, PhysRevD.101.063025,https://doi.org/10.48550/arxiv.2211.12808}. 

\section{Summary and Conclusion}
\label{summary}

In this work, we studied the radial oscillations of $\Delta$-inclusive neutron and hyperon stars
 employing the DD-RMF model with the DD-ME2 parameter set. The spin-$3/2$ baryons ($\Delta$s) are described using the Rarita-Schwinger Lagrangian density. For the spin-3/2 decuplet and the spin-1/2 baryonic octet, the baryon-meson coupling constants are calculated using the Clebsch-Gordan coefficients of the SU(3) group. The coupling constants of the scalar meson are fixed to replicate the known potential depth using a QHD model that essentially satisfies all requirements at the saturation density, thus allowing a unified approach to the coupling constants of hyperons and delta resonances. We studied the 20 lowest eigenfrequencies and corresponding oscillation functions of $\Delta$-inclusive nuclear (N+$\Delta$) and hyperonic matter (N+H+$\Delta$) by solving the Sturm-Liouville boundary value problem and also verifying its validity.
 For the hydrostatic equilibrium, we numerically solved the structural equations to obtain the mass-radius relationship of $\Delta$-inclusive neutron and hyperon stars. The Sturm-Liouville equations were then solved for the perturbations imposing the necessary boundary conditions in order to examine radial oscillations of pulsating stars. This allowed us to calculate the frequencies of the modes as well as the related wave functions. 19 excited $p$-modes and the fundamental $f$-mode have been calculated. The addition of hyperons softens the EoS, decreasing the maximum mass and hence increasing the corresponding frequencies of the pulsating star. While the addition of $\Delta$ baryons to nucleonic matter softens the EoS, it gets stiffer for the hyperonic matter. Compared to hyperonic matter, the adiabatic index $\gamma$ for the $\Delta$-inclusive matter exhibits far more complex behavior. Due to the onset of the $\Delta^-$, we observe a significant decrease in the value of the parameter followed by a rapid increase. This increases even more at intermediate densities than it does for the pure nucleonic case, a behavior not seen in the hyperonic case.

 We investigated the radial displacement perturbation profile $\xi(r)$ and pressure perturbation profile $\eta(r)$ with $\Delta$-inclusive matter and found that they oscillate with exactly $n$ nodes for the $n^{th}$ mode for all cases. The lowest modes show a smooth drop in their profiles while the higher modes depict lower oscillations which become large for higher modes. For $\Delta$-inclusive nucleonic matter, small kinks are present for $n$ = 0 mode. For hyperonic matter, the kink at the same node is found to be large and present at a small radius. These kinks in $\xi(r)$ correspond to the emergence of new exotic particles which provides a discontinuity in the adiabatic index $\gamma$.
 We see that the lowest mode frequencies for N+$\Delta$ and N+H EoSs are higher as compared to the pure nucleonic matter because of the deltas and hyperons. Furthermore, the separation between consecutive modes increases with the addition of hyperons and $\Delta$s.

One of the main reasons for an abrupt change in the oscillatory property of the star as one undertakes a small variation in the star's stellar configuration is the precisely defined division of the star's inner core and outer crust. These two regions have different EoS and hence different oscillation properties. The stellar configurations such as the star's mass or central energy density essentially determine the oscillation frequencies of the crust and core pulsations. For the crust part, while the oscillation functions are expanding pretty steeply, they either decline or essentially remain constant for the NS core.

The oscillatory properties of the star could be due to pulsations in the core or crust of the NS. The mass of the star affects the frequency spectrum. For moderately massive NSs, the oscillatory properties of the lower-order radial modes are determined by the core pulsations. If the stellar mass is low enough or the frequency is high enough, the star may be affected by the crust pulsations. The avoided crossing phenomena are closely related to the changes in matter's compressibility all across the star, which are modeled by the adiabatic index $\gamma$. The maximum of $\gamma$ occurs close to the boundary between the core and crust as the stiffness of the matter increases outward.

Unlike non-radial oscillations, radial oscillations do not possess a gravitational wave counterpart, making them a distinctive means to directly discern the influence of the EoS on the structure of the neutron star. This allows for a more straightforward analysis without the added complexities associated with gravitational wave measurements. 

Observing multiple radial modes, including the fundamental mode ($f$-mode) and pressure modes ($p$-modes), offers a precise means of measuring the radius of compact stars. This methodology has proven successful in other branches of Asteroseismology, demonstrating its reliability and applicability.

The future observation of multiple radial oscillation modes, particularly through the computation of the large separation, holds the potential to identify the presence of delta baryons or hyperons within neutron stars. This
would contribute to validating the existence or absence of these species in different regions of the NS.

As a result, it is possible to probe the exotic degrees of freedom existing inside the NS using the high sensitivity of the enormous separation to the interior structure of the star. The properties of more realistic environments, such as temperature, rotation, and magnetic field, should also be incorporated in order to study radial oscillations in newborn NSs following supernova explosions or the merger of NSs. We leave these studies for future work.

\section{Acknowledgement}
We would like to thank the anonymous reviewer for valuable comments and suggestions.
I. A. R. and I. L. acknowledge the Funda\c c\~ao para a Ci\^encia e Tecnologia (FCT), Portugal,
for the financial support to the Center for Astrophysics and Gravitation (CENTRA/IST/ULisboa)
through the grant Project~No.~UIDB/00099/2020  and grant No. PTDC/FIS-AST/28920/2017.
K. D. M. was supported by the Conselho Nacional de Desenvolvimento Científico e Tecnológico (CNPq/Brazil) under grant 150751/2022-2.

%
\end{document}